\documentclass[twocolumn,           
               showpacs,            
               noshowkeys,
               preprintnumbers,     
               aps,                 
               prd,          	    
               letterpaper,             
               superscriptaddress,      
               nofootinbib,         
               tightenlines,        
               floats,floatfix      
               ]{revtex4-1}

\usepackage{graphicx}
\usepackage{dcolumn}
\usepackage{bm}
\usepackage{amsmath}
\usepackage{amsfonts}

\def\bx{{\bf x}}

\newcommand{\deltath}{\delta_{\rm th}}

\newcommand{\betapbhs}{\beta_{\rm PBH}}
\newcommand{\beq}{\begin{equation}}
\newcommand{\eeq}{\end{equation}}
\newcommand{\bea}{\begin{eqnarray}}
\newcommand{\eea}{\end{eqnarray}} 
\newcommand{\eref}[1]{Eq.~\eqref{#1}}

\begin{document}

\title{Formation of subhorizon black holes from preheating}

\author{Efra\'in Torres-Lomas}
\email[]{efrain@fisica.ugto.mx}
\altaffiliation{}
\affiliation{%
Departamento de F\'isica, Divisi\'on de Ciencias e Ingenier\'ias, Campus Le\'on, 
Universidad de Guanajuato, Le\'on 37150, Guanajuato, M\'exico}

\author{Juan Carlos Hidalgo}
\email[]{hidalgo@fis.unam.mx}
\affiliation{Institute of Cosmology and Gravitation, University of Portsmouth, Dennis Sciama Building, Portsmouth PO1 3FX, United Kingdom.}
\affiliation{Instituto de Ciencias F\'isicas, Universidad Nacional
  Aut\'onoma de M\'exico. Apdo. Postal 48–3, 62251, Cuernavaca,
  Morelos, M\'exico.}

\author{Karim A. Malik}
\email[]{k.malik@qmul.ac.uk}
\affiliation{Astronomy Unit, School of Physics and Astronomy,
Queen Mary University of London,Mile End Road, London, E1 4NS, United Kingdom}

\author{L. Arturo Ure\~na-L\'opez}
\email[]{lurena@fisica.ugto.mx}
\affiliation{%
Departamento de F\'isica, Divisi\'on de Ciencias e Ingenier\'ias, Campus Le\'on, 
Universidad de Guanajuato, Le\'on 37150, Guanajuato, M\'exico}

\date{\today}

\begin{abstract}

We study the production of primordial black holes (PBHs) during the
preheating stage that follows a chaotic inflationary phase. The scalar
fields present in the process are evolved numerically using a modified
version of the \textsc{HLATTICE} code. From the output of the numerical
simulation, we compute the probability distribution of curvature
fluctuations, paying particular attention to sub-horizon scales. We
find that in some specific models these modes grow to large amplitudes
developing highly non-Gaussian probability distributions. We then
calculate PBH abundances using the standard Press-Schechter criterion
and find that overproduction of PBHs is likely in some regions of the
chaotic preheating parameter space.
\end{abstract}



\maketitle

\section{Introduction}
\label{sec:1}

During primordial inflation, spacetime expands exponentially for about
60 $e$-folds, producing the homogeneous, isotropic and almost flat
Universe we observe today. Small fluctuations of the inflationary
field are stretched to scales larger than the cosmological horizon and
reenter in subsequent epochs to source the cosmic structures. After
inflation ends, the energy stored in the dominant field must decay
into relativistic particles to create the radiation-dominated
environment required by nucleosynthesis, a transition phase referred to
as reheating. Modelling reheating remains a challenge since one must deal with the evolution
of highly inhomogeneous fields in an expanding background, including
nonlinear phenomena up to the time of thermalization of the Universe.
  
Of particular interest in recent years has been a
reheating model in which resonant amplification by the inflaton field
leads to particle production, a process known as preheating
\cite{Traschen:1990sw,*Shtanov:1994ce,*Kofman:1994rk}.
Here a spectator field $\chi$ is nonminimally coupled with the
dominant inflaton $\phi$, which oscillates at the bottom of the
potential. The quantum fluctuations of $\chi$ experience a resonant
amplification, causing an exponential growth of its occupation numbers
and an explosive production of relativistic particles. The
parametric-resonant mechanism of preheating has proved to be extremely
efficient for a range of parameter values \cite{Kofman:1997yn}.    
To improve our understanding of preheating models we can study the
gravitational instability that will amplify the matter
fluctuations. Large matter overdensities may form due to the highly
nonlinear physics of the preheating mechanism 
\cite{Green:2000he,Amin:2011hj}, and the highest density
concentrations may collapse and form black holes. 
Consequently, the observable constraints on the abundance of these
primordial black holes (PBHs) could help us to constrain models of
preheating. The goal of this paper is to determine the production
rate of PBHs during the preheating after a chaotic inflation stage. 
In this preheating model, the inflaton  couples nonminimally to
 a massless spectator field through a ``four-legs interaction'' term.
The evolution of the scalar fields is calculated numerically with a
simplified version of the \textsc{HLATTICE} code \cite{Huang:2011gf}, which
performs a three-dimensional integration of the equations of motion for the full
nonlinear variables. We find that the amplitude of fluctuations
increases rapidly inside the Hubble horizon even when the background
energy component behaves almost like radiation. At the same time, the
probability distribution of curvature inhomogeneities develops a
skewed profile. We provide an example where both effects conspire to
produce a considerable number of PBHs at subhorizon scales (a
possibility explored in Ref.~\cite{Lyth:2005ze}).

Our paper is organized as follows. The following section presents the
elements of the preheating model we work with, including a description
of the numerical setup in the present study. Section~\ref{sec:3}
describes the criteria used to account for the formation of PBHs from
overdensities at subhorizon scales in the preheating phase. In
Sec. \ref{sec:4} we present the probability density distribution for
inhomogeneities in our model and estimate the probability of PBH
formation. We conclude in Sec. \ref{sec:5} with a summary and a
discussion of the possible extensions to our study. 

\section{Preheating model}
\label{sec:2}

The preheating model we consider is given by the Lagrangian 
\begin{equation}
\label{chaotic:model}
  {\cal L} = \frac{1}{2} \phi_{,\alpha} \phi^{,\alpha}  +
  \frac{1}{2} \chi_{,\alpha}\chi^{,\alpha}  - \frac{1}{2} m^2\phi^2 -
  \frac{1}{2} g^2 \phi^2 \chi^2 \, ,  
\end{equation}
\noindent where $\phi(\bx,t)$ is the inflaton, and
$\chi(\bx,t)$ is the auxiliary, spectator field;
both are non-linear functions of time and space. The index $\alpha$
denotes the spacetime coordinates $(0,1,2,3)$, $m=10^{-6} m_{\rm Pl}$
is the inflaton mass, and $g^2$ is the coupling constant, which is a free
parameter in our study.  

To study the evolution of the matter fields together with that of the
spacetime, we have used a modified version of the \textsc{HLATTICE} code
\cite{Huang:2011gf}, with a flat
Friedmann\-Lemaitre\-Robertson\-Walker metric, for which the dynamics is governed by
the homogeneous scale factor $a(t)$ only. 
Note that this implies the suppression of spacetime fluctuations in
our simulations, but it can be regarded as the choice of a flat
gauge in the context of cosmological perturbation theory
\cite{Malik:09}. Each hypersurface of constant time is thus conformal
to flat space (we shall discuss this point further in 
Sec.~\ref{sec:3}).

Our numerical simulation starts from about one $e$-fold before the end of
inflation (defined here as the time when the Hubble scale finds a
global minimum). We choose background values for the fields at the initial
time as $\phi_{\rm init} = \bar{\phi} = 0.3 M_{\rm Pl}$ and
$\chi_{\rm init} = \bar{\chi} = 0$, with a Gaussian
distribution of perturbations around these mean values, defined in
Fourier space as $|f_k|^2 = 1/(2 \omega_k)$ and
$|\dot{f}_k|^2=\omega_k/2$. Here $\omega_k = \sqrt{k^2+m^2_f}$ for each
one of the fields $f:(\phi,\chi)$. From the Lagrangian in
\eref{chaotic:model} we can read the squared effective masses $m^2_f$ of the fields
as $m^2_\phi = m^2 + g^2 \chi^2$ and $m^2_\chi = g^2 \phi^2$.
In particular we will test for PBH formation considering two values of $g^2$ 
commonly used in preheating studies: 
\emph{\bf case} I takes $g^2=6.5 \times 10^{-8}$, and in \emph{\bf case II}, we
consider $g^2 = 2.5\times10^{-7}$ (more details of the simulations can
be found in Refs.~\cite{Huang:2011gf,Felder:2000hq,Torres-Lomas:2012tna}).   

The growth of the modes $\chi_k$ in Fourier space yields an increase
in the number of created particles. 
Indeed, the number density $n_k$ of particles with momentum ${\bf k}$
can be evaluated as the energy of that mode divided by the energy of 
each particle, obtaining  
\begin{equation}
\label{occ:number}
n_k={\omega_k\over 2} \left( { |\dot \chi_k|^2 \over \omega^2_k}
+|\chi_k|^2 \right)-{1\over 2},
\end{equation} 
and, following Ref.~\cite{Huang:2011gf}, we approximate  $\omega_k \simeq k$
to avoid ambiguities in the nonlinear regime.

\begin{figure}[tbp]
\includegraphics[width=0.5\textwidth,height=0.25\textheight]{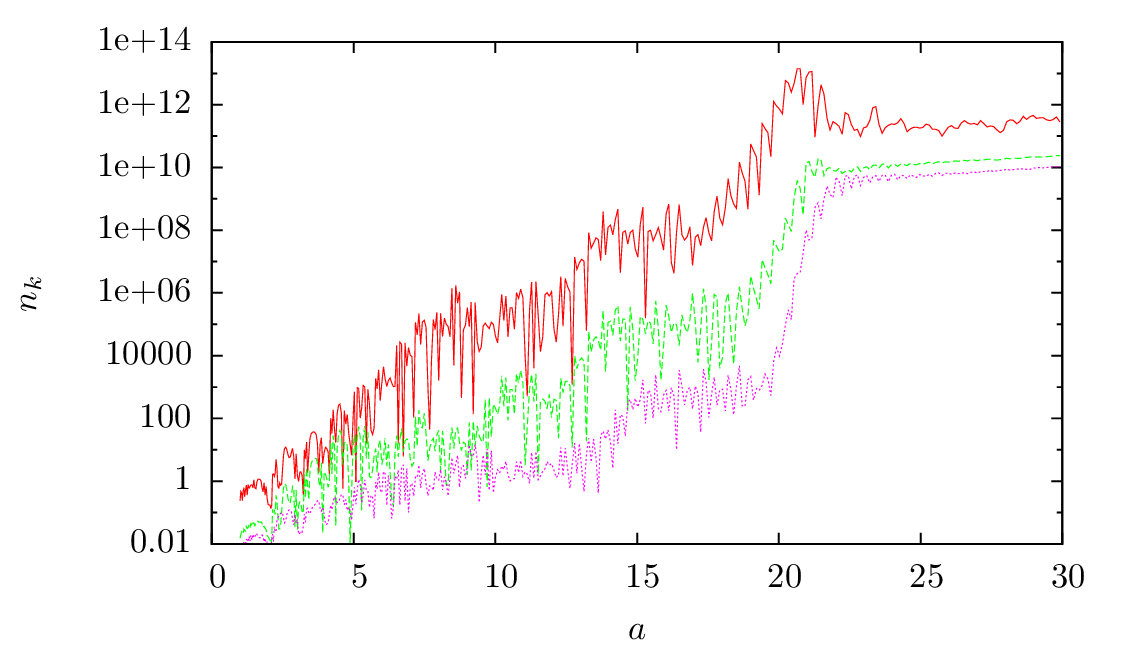}
\caption{
Time evolution of occupation number $n_k$ as defined in Eq.~\eqref{occ:number}. 
The large value of $n_k$ justifies the classic treatment of inhomogeneities in
our problem. The top line represent a mode that started outside of the
horizon, and middle and lower lines refer to modes that remained inside
the horizon at the end of inflation. 
 }  
\label{fig:1}
\end{figure}

Figure~\ref{fig:1} shows the exponential
increase in $n_k$ for modes in the resonance band
with the characteristic steplike profile observed in previous works
(e.g. Ref.\cite{Kofman:1997yn}). We confirm the exponential growth of the
occupation number but in an dynamic background, an improvement over
Ref.~\cite{Kofman:1997yn}, which follows the evolution of scalar field in a
spacetime dominated by dust (c.f.~Fig.~4 of
Ref.~\cite{Kofman:1997yn}). As indicated
in Ref.~\cite{Podolsky:2005bw}, $n_k$ helps us to understand the energy
distribution for each mode. Once the 
distribution hits the highest $k$ mode resolved by the simulation,
the energy is reflected back to the infrared modes.
For the model and resolution considered here, the
numerical simulation can be trusted up to a final time $t_{\rm f}$ where $a(t_{\rm
  f}) = 30$ (well within the broad parametric resonance stage), just
before the energy reflection effect kicks in and develops spurious
modes. 

Because of the exponential growth in the occupation number
shown in Fig.~\ref{fig:1},
we can treat the energy density 
in classical terms nearly after the end of inflation. Thus, we are 
free to study the formation of PBHs following the usual methods. 
The total energy density and pressure are defined by
\begin{eqnarray}
\label{total:density}
\rho(\bx,t) &=& \frac{1}{2} \dot{\phi}^2 + \frac{1}{2} (\nabla \phi )^2 +
\frac{1}{2} \dot{\chi}^2 + \frac{1}{2} (\nabla \chi )^2 + V(\phi,\chi ) \\
p(\bx,t) &=& \frac{1}{2} \dot{\phi}^2 - \frac{1}{6} (\nabla \phi )^2 +
\frac{1}{2} \dot{\chi}^2 - \frac{1}{6} (\nabla \chi )^2 - V(\phi,\chi ),
\notag 
\end{eqnarray}

\noindent and can be split in a homogeneous part and inhomogeneous
fluctuations. The homogeneous part of the energy density and pressure,
$\bar{\rho}(t)$ and $\bar{p}(t)$, are defined through their spatial
average at every time step during the simulations.  
Instead of plotting the behavior of these homogeneous quantities, we
follow the evolution of the time dependent equation of state $w(t) \equiv \bar{p} / \bar{\rho}$,
which plays a significant role on both the analysis of
matter-radiation transition and in determining the criterion of gravitational
collapse of overdense regions into PBHs.

The oscillations of the inflaton about the minimum of its quadratic
potential are translated into oscillations of $w$, with time average
$\langle w \rangle \approx 0$ at the beginning, and, by the energy
transference to the (massless) spectator  field,  $0< \langle w
\rangle \lesssim 1/3$ at the end of the simulation, Note that the
complete radiation domination can be achieved only in an ideal
preheating model. 

To transfer energy more effectively and to produce higher redispersion effects between the $k$ modes of both fields, which eventually break the oscillations of the background quantities and stabilize the equation of state, we could select a higher value for the coupling constant $g^2$. But, due to the high nonlinearity of the process, that naive selection does not necessarily improve the preheating process.

\begin{figure} [t]
\includegraphics[width=0.5\textwidth]{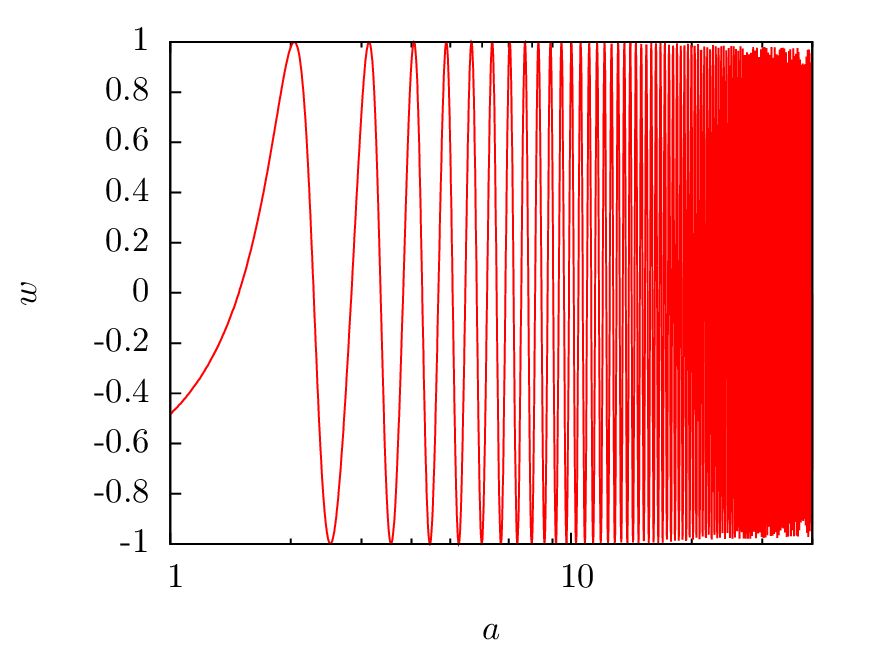}
\includegraphics[width=0.5\textwidth]{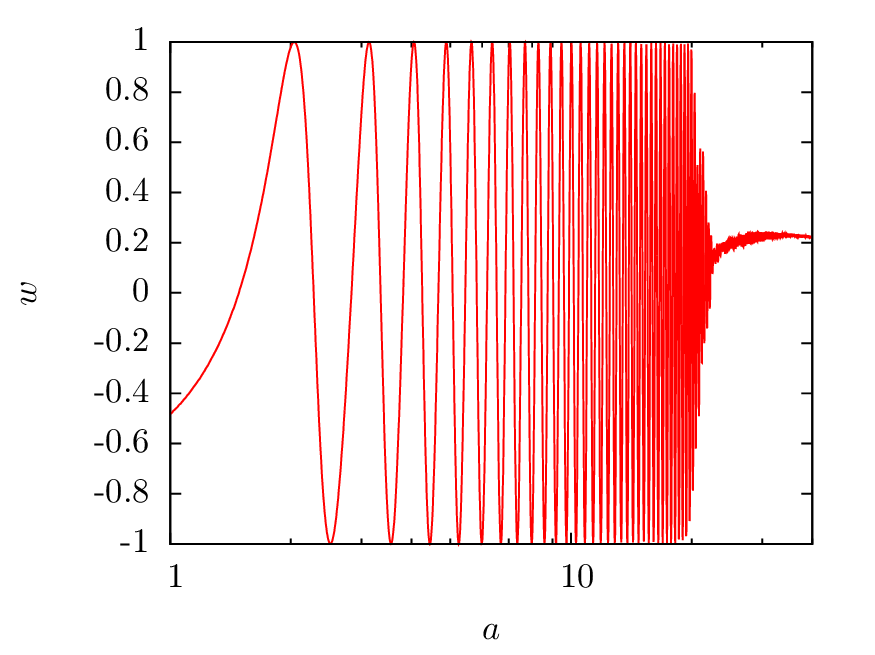}
\caption{Time evolution of the equation of state $w$ for case I (top) and
case II (bottom).Only for the latter $w$ steadily converges
  toward a constant value. See the text for more details. } 
\label{fig:2}
\end{figure}

Figure~\ref{fig:2} shows that the stabilization of $w$ at
the end of preheating is not generic in the model \eqref{chaotic:model}, and neither
case I nor Case II are able to produce an ideal complete preheating, even though case II produces a  $w\approx cte=0.2$
(more general analyses about the effectiveness of the model
\eqref{chaotic:model} are reported in Ref.~\cite{Podolsky:2005bw}).

\section{PBH formation criteria}
\label{sec:3}


Primordial black holes form from overdense regions in high density
environments. Fluctuations with large density
contrast  $\delta  = \delta \rho /\bar{\rho}$ overcome the 
pressure of the environment and detach from the background
expansion. In the standard scenario,if the amplitude of an overdensity of size 
$L$ is large enough, it will recollapse as soon as it enters in
causal contact, that is, when
$L_H = 1 / H $.  As a result of the gravitational collapse, a black
hole of mass comparable to the Hubble mass is formed right after horizon
crossing. The threshold amplitude for the density contrast $\deltath$
to reach the collapse was first estimated to be $\deltath \simeq w$
 in a barotropic fluid with pressure $p = w \rho$ using the
   Jeans criterion for overdensities in a Friedmann background
   \cite{Carr:1975qj}.  
Recently, simulations considering inhomogeneous cosmologies have
determined a threshold amplitude for the comoving matter overdensity
as $\deltath \approx 0.41$ in a pure radiation background
\cite{Musco:2004ak}. Extensions to general barotropic fluid
backgrounds are considered in Ref.~\cite{Musco:2012au,*Harada:2013epa}.  

The threshold for collapse is best expressed in terms of the
gauge-invariant curvature perturbation $\zeta$
\cite{Shibata:1999zs,*Green:2004wb}. This is defined in the uniform
density gauge as  
\beq
\label{zeta:def}
\zeta = -\psi - H \frac{\delta \rho }{\dot{\bar{\rho}}}.
\eeq

\begin{figure}[t]
\includegraphics[width=0.5\textwidth,height=0.23\textheight]{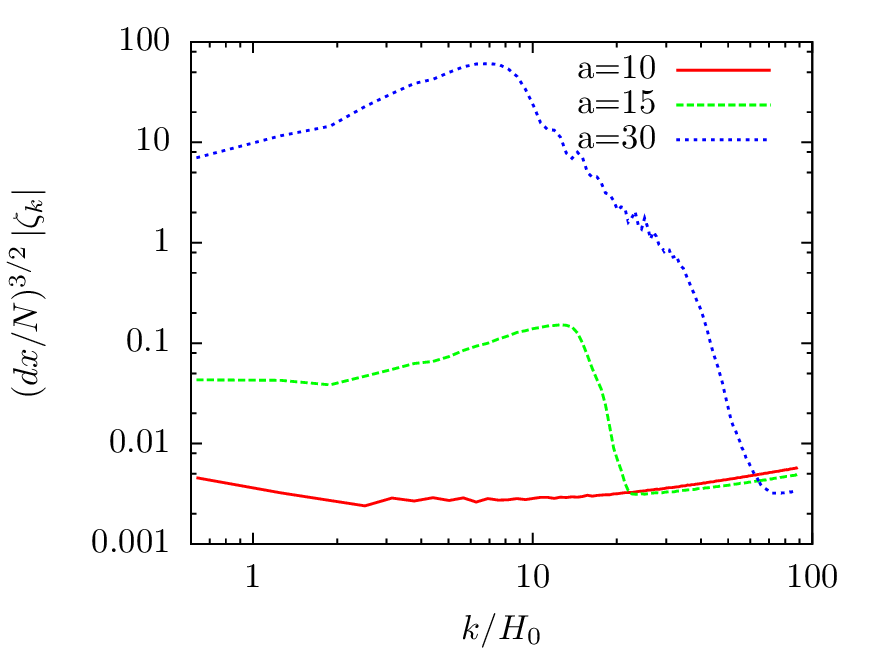}
\includegraphics[width=0.5\textwidth,height=0.23\textheight]{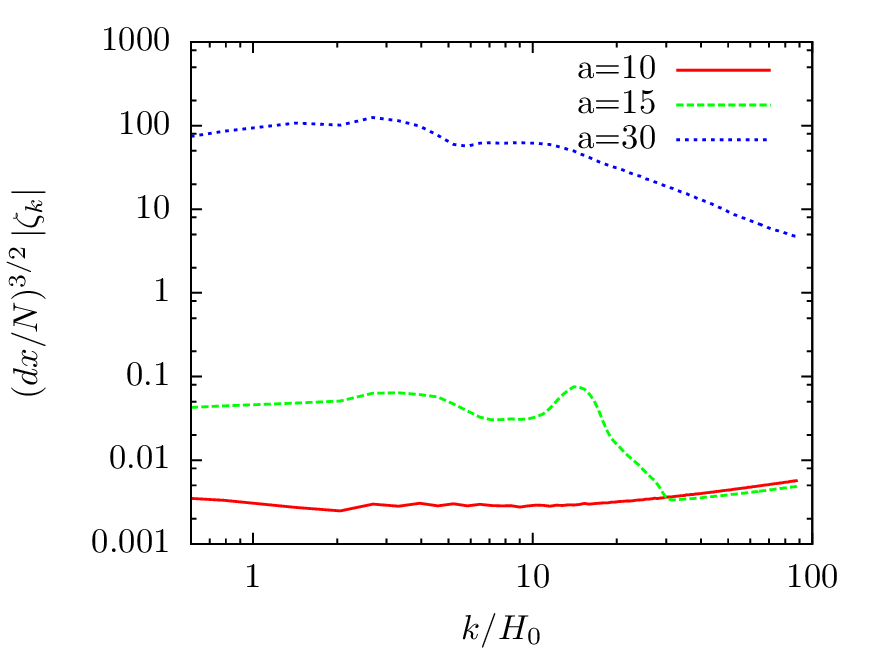}
\caption{Time evolution of the mean amplitude of
  $\zeta_k(t)$ as a function of the wave number $k$. Note the
  development of a maximum for a particular scale is formed inside the
  initial cosmological horizon scale $L_{H0} = 1 / H_0$.}  
\label{fig:3}
\end{figure}

For the cases that concern us, we note from Fig.~\ref{fig:3} that
during preheating the fluctuations grow in amplitude inside the
cosmological horizon (as we shall see in detail in Sec.
\ref{sec:4}). This indicates that PBHs are more likely to form inside
the cosmological horizon, and a different criterion is required to set
the threshold of amplitudes that collapse and form PBHs.  The
formation of PBHs at scales inside the Hubble horizon has been largely
ignored due to the linear Jeans instability criterion, which for a
radiation-dominated universe sets the scale of instability close to
the Hubble scale. Furthermore, in the dustlike environment, typical
of phase transitions in unification theories, PBHs are thought to form
at small scales and from overdensities of arbitrarily small
amplitude because there is no pressure to prevent the collapse
\cite{Khlopov:1980mg}. However, scattering of small black holes could
prevent the formation of larger PBHs, e.g.,~Ref.~\cite{Alabidi:2013lya}.
In Ref.~\cite{Lyth:2005ze}, the formation of PBHs at subhorizon scales
is studied. The authors show that the threshold amplitude for the
metric fluctuation to form a black hole at scales well inside the
horizon can be taken to be equivalent to the known $\zeta_{\rm th} =
0.7$ of the radiation background.\footnote{The threshold value given in Ref.~\cite{Lyth:2005ze} (see also Ref.~\cite{Zaballa:2006kh}) and used in this work is a conservative estimate. The precise determination of the sub-horizon threshold
value is beyond the scope of this paper.} Starting from the matter
overdensity in the flat gauge $\delta_{\rm flat}$, we can compute the
curvature fluctuation as
\eref{zeta:def}, 
\beq
\label{zeta:threshold}
\zeta = - H \frac{\delta \rho_{\rm flat} }{\dot{\bar{\rho}}} =
\frac13 \frac{\deltath}{1 + w }.
\eeq
\noindent   

A key result in our simulation is presented in Fig.~\ref{fig:3},
where we plot the evolution of the
modes $\zeta_k(t)$  over the range of comoving scales covered by the
simulation box; more precisely, we plot the evolution of the combination
 $(dx/N)^{3/2} \left | \zeta_k \right | $, which is a box size independent quantity. 
 This results show that the curvature perturbation grows and tends to peak at 
 scales below the Hubble horizon; this motivates the consideration of PBH formation at subhorizon scales.   

In the following, we use the threshold value $\zeta_{\rm th} = 0.7$ to
select those overdensities that will inevitably form black holes at
subhorizon scales and look into the abundance of PBHs. 

It is important to stress that all the analytical and numerical
threshold estimates have been  
found for cosmological phases characterized by a \textit{constant} $w$, and this 
validates our threshold definition at the end of preheating for case
II, where $w$ is stabilized by the redispersion effects. For  case I,
the validity of $\zeta_{\rm th} = 0.7$ is less clear and must be
considered with some care.

\section{Probability of PBH formation in preheating}
\label{sec:4}

\begin{figure}[tb]
\includegraphics[width=0.5\textwidth,height=0.24\textheight]{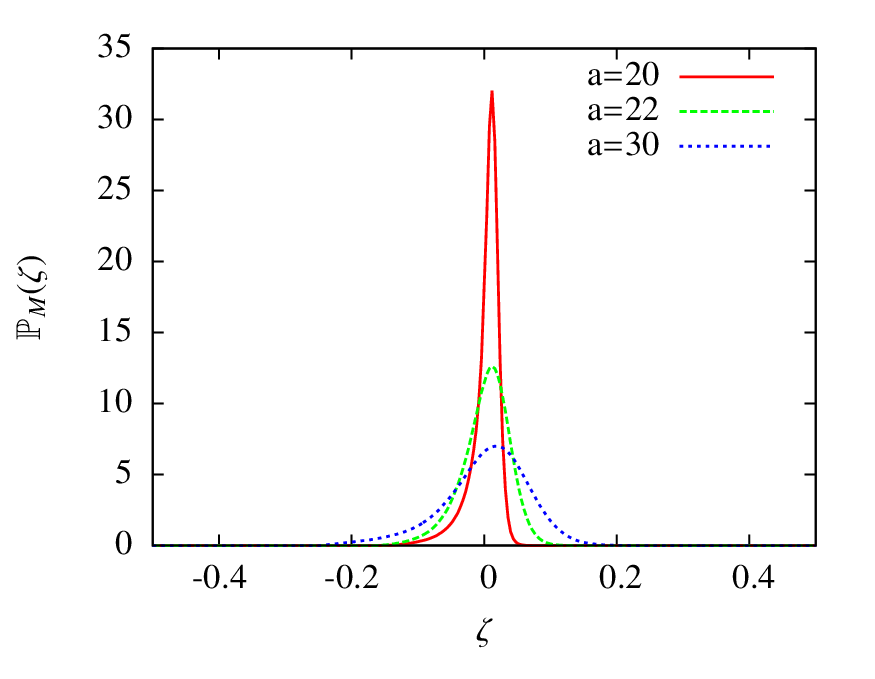}
\includegraphics[width=0.5\textwidth,height=0.24\textheight]{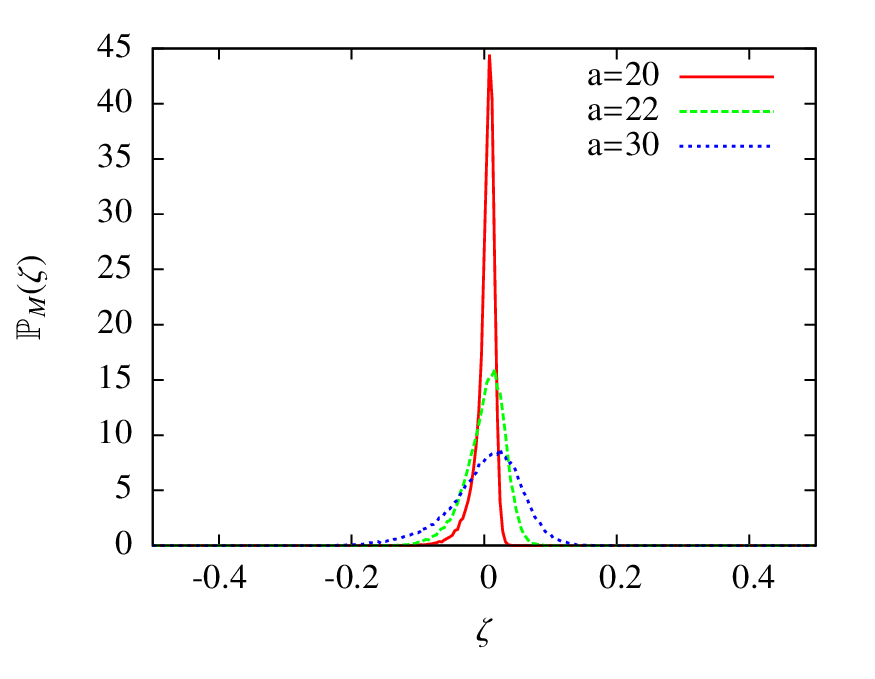}
\caption{Probability distribution function of $\zeta_k(t)$ for two modes at
  different times of evolution in case I.
  The top (bottom) panel represents the PDF for a mode $ L_H(t_{\rm
    f}) /L = 46$  ($L_H(t_{\rm f}) /L  = 23$) at the end of the 
  simulation  when $a(t_{f}) = 30$.  Note that the shape of the
  distribution evolves with time through the preheating phase
  generating negatively skewed distributions.}  
\label{fig:4}
\end{figure}

\begin{figure}[tb]
\includegraphics[width=0.5\textwidth,height=0.24\textheight]{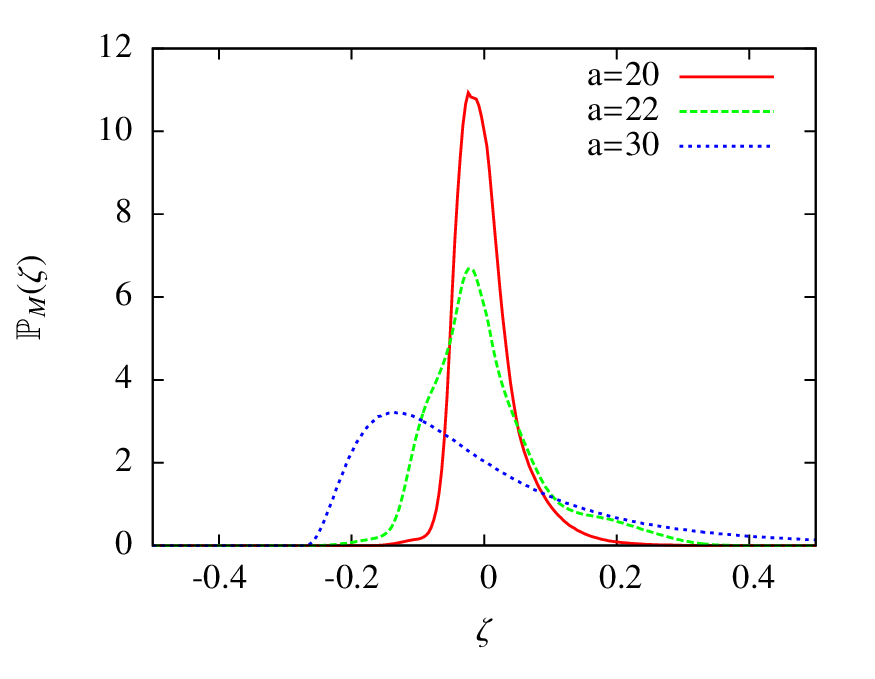}
\includegraphics[width=0.5\textwidth,height=0.24\textheight]{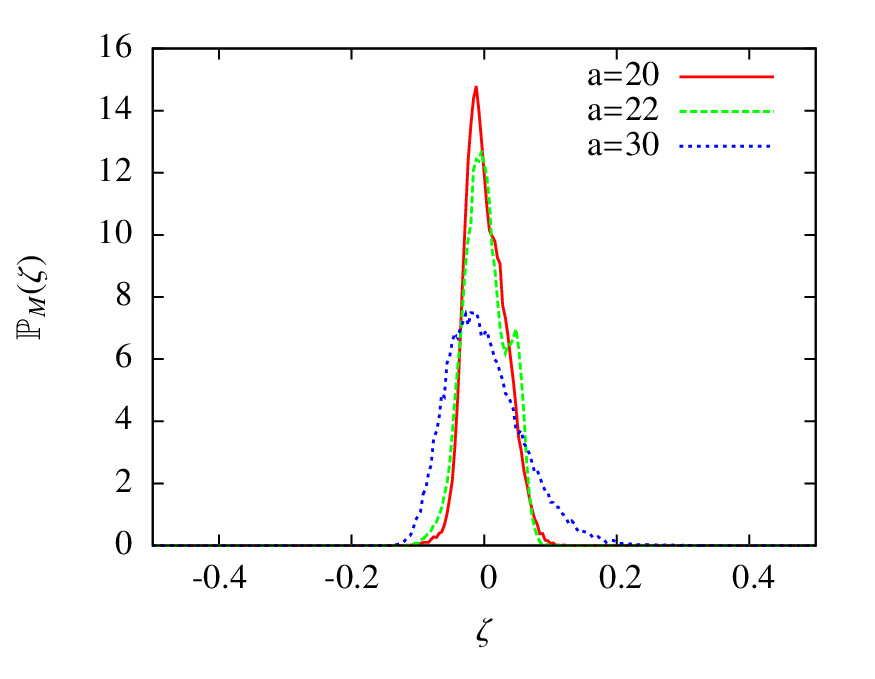}
\caption{Same as Fig.~\ref{fig:4} but for the model of case II.
  In this case, the distributions are positively skewed, a feature that
  favors the formation of PBHs.}   
\label{fig:5}
\end{figure}

To estimate the mass fraction of black holes at $t= t_{\rm f}$,
$\betapbhs(M) \equiv \rho_{\rm PBH} / \rho(t_{\rm f}) \simeq \rho_{\rm PBH} / \bar{\rho}(t_{\rm f})$, we first construct
the probability distribution of the curvature perturbation $\zeta$ at
a specific scale $k$ [or the equivalent mass $M \propto
\bar\rho(t_{\rm f}) k^{-3}$], namely, $\mathbb{P}_M(\zeta)$. As shown in
Figs.~\ref{fig:4} and \ref{fig:5}, the probability distribution function (PDF)
develops an appreciable skewness below the Hubble scale
$L_H$.

In general, a non-Gaussian distribution generates nonvanishing
moments beyond the variance in real space; likewise, in Fourier space, it generates nonvanishing correlation functions beyond the two-point function. For example, in a typical cosmic microwave background (CMB) anisotropy analysis, the contribution of the reduced bispectrum $f_{\rm NL}(k_1,k_2,k_3)$ \cite{Lyth:2009zz} denotes the amplitude of the three-point function. The explicit contribution of $f_{\rm NL}$ to the abundance of PBHs $\beta_{PBH}$ has been considered by several authors
\cite{Hidalgo:2007vk,Saito:2008em,Byrnes:2012yx,1475-7516-2013-08-052}, showing that the bispectrum can significantly modify the abundance of PBHs with respect to the Gaussian case. This may be used to constrain
the non-Gaussian parameters, $f_{\rm NL}$, $g_{\rm NL}$, etc., from
the observational bounds on the abundance of PBHs. However, a PDF reconstructed from the cosmological non-Gaussianity parameters alone could fail to reproduce the profile of the distribution tail~\cite{Shandera:2012ke}, which precisely accounts for the large amplitudes we want to integrate. Because of this, and since we know the solutions in real space, we focus our attention on the PDF that directly arises from the distribution of $\zeta({\bf x})$ in the lattice.

As customarily done, we use the Press\-Schechter formalism to
compute the mass fraction from the non-Gaussian PDF. Applied to the
distribution of curvature fluctuations, this yields 
\beq
  \label{mass:fraction}
  \betapbhs(M) 
  = 2 \int_{\zeta_{\rm th}}^{\infty} \mathbb{P}_M(\zeta) d\zeta\,.
\eeq
%

We have computed density profiles  $\mathbb{P}(\zeta)$ for different scales by
looking at the overdensities of size a few times the resolution scale. In the
plots of Fig.~\ref{fig:4} for the case~I, the mean amplitude of $\zeta_k$
shows a negatively skewed distribution for modes inside the horizon, while
for case II, the distribution is positively skewed. This leads to an
enhancement in PBH formation (with respect to the Gaussian) for the
model of case II.

\begin{figure}[h]
\includegraphics[width=0.45\textwidth]{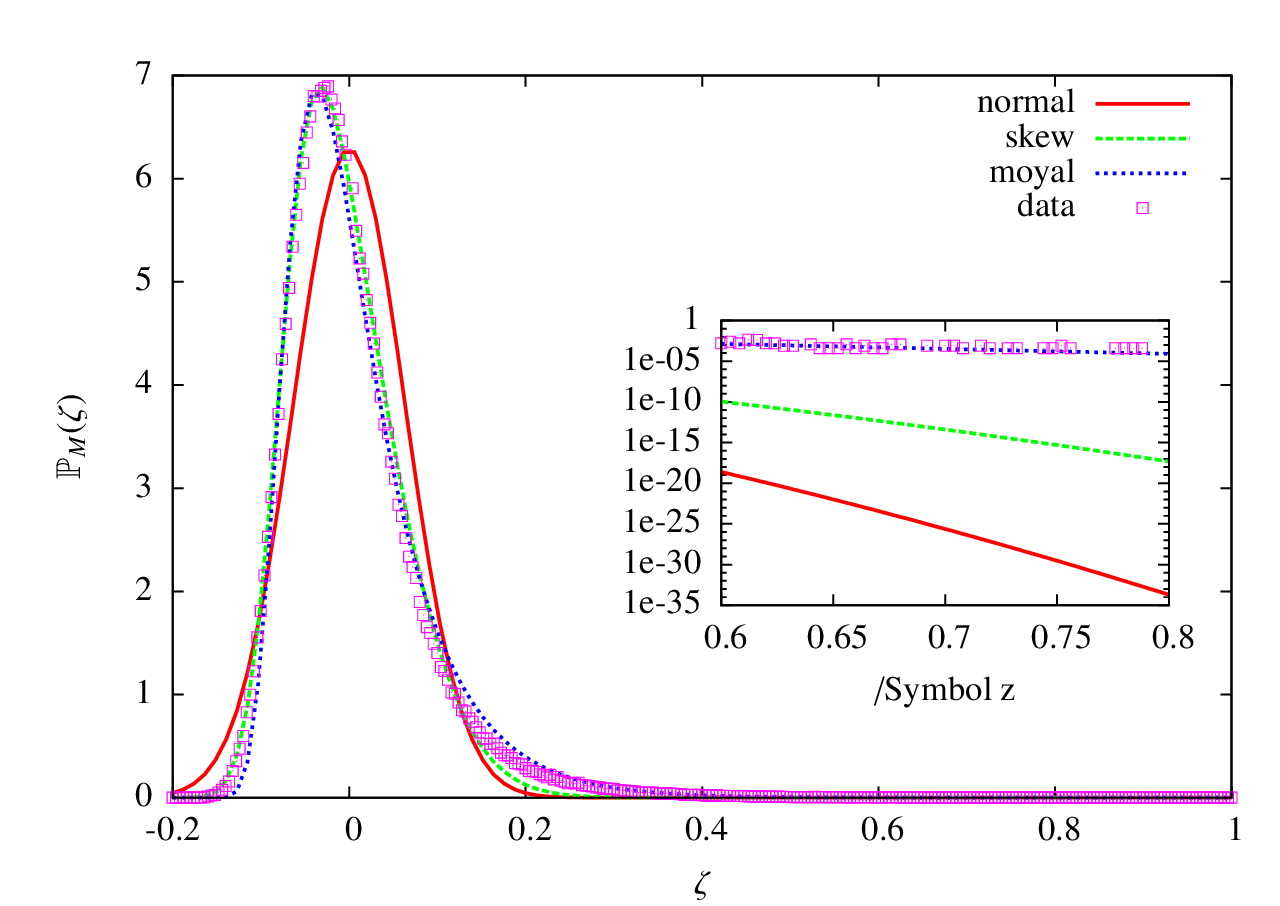}
\caption{Tail of the PDF for the scale
  $L_H(t_{\rm f}) /L  = 23$.  The true normalized distribution is approximated by two
  analytical PDFs. From below, the skewed normal distribution, and from above, the closest  approximation is a Moyal distribution, which is the best
  approximation to the tail of the distribution.}
\label{fig:6}
\end{figure}

Because of the limited size of our simulation box, the incomplete sample
of modes translates into a truncated tail of the numerically generated PDF. To
estimate the complete probability of large amplitude fluctuations, we
carefully fitted the numerical PDF with analytic distributions which
bound the true distribution from above and below (see
Fig.~\ref{fig:6}). For these analytical PDFs, we can  
easily evaluate the integral of \eref{mass:fraction}. As a result, we
compute upper and lower bounds to the true $\betapbhs$. In
Table~\ref{table:1}, we present approximate values of $\betapbhs$, calculated 
at $t_f$, for overdensities  of size of a fraction $L/L_H$ of the
horizon scale.   
The value of $\betapbhs$ increases with the wave number as a 
consequence of the non-Gaussian evolution of fluctuations inside the 
cosmological horizon. The development of non-Gaussianity of
the PDFs is evident from the snapshots of evolution illustrated in
Figs.~\ref{fig:4}~and \ref{fig:5}. 

Let us finish this section by analyzing the results of Table~\ref{table:1}
in light of the bounds to the PBH mass fraction at the smallest
scales. It is known that the Hawking evaporation process of PBHs may
halt at the Planck scale. The remnant Planck-mass black holes would
survive behaving as a component of the dark matter until the present
time. Since their mass fraction cannot exceed that of dark matter, one
can impose a bound to their abundance, that is \cite{Carr:1994ar},
\begin{equation}
\beta_{\rm  PBH}(M_{\rm PL}) < 10^{-28}\left(\frac{M}{M_{\rm Pl}}\right)^{3/2}
\label{planck:bound}
\end{equation} 

\noindent where $M_{\rm Pl}$ is the Planck mass. When we compute the physical mass
enclosed in a region of radius $L / L_H = 1/16$ at $t_{\rm f}$, we note
that the PBHs formed at that scale would have $M_{\rm PBH} \approx
10^6 M_{\rm Pl}$. For this scale, we note that the minimum value $\beta_{\rm
  min}$ reported in Table~\ref{table:1} exceeds the bound
imposed by Eq.~\eqref{planck:bound}. On larger scales, $\betapbhs$
falls below the observational bounds. As for smaller scales, they lie too
close to the resolution scale of our simulation, and thus we do not
consider them physical (a complete set of bounds is
reported in Ref.~\cite{Carr:2009jm} and see also Ref.~\cite{Josan:2009qn} for
specific bounds to $\zeta_k$). The implications and limitations of the
results derived are discussed in the following last section.  
   
\section{Discussion}
\label{sec:5}

In this paper, we have considered the possibility of
forming sub-Hubble scale black holes, solving numerically the classical
governing equations of a model of preheating after chaotic inflation.
We have focused on computing the distribution of the curvature
fluctuation $\zeta$ from the sample of modes in the simulation box
of size smaller than the Hubble scale $L_{H}$. For the typical values of
the coupling parameter $g^2 = 2.5 \times 10^{-7}$ in the reheating
model (case~II of this paper), we find that the small size
inhomogeneities grow in amplitude. When constructing the PDF, we note that large skewness is developed in the 
probability profile as the modes enter the cosmological horizon
(cf. Fig.~\ref{fig:5}). We find
that, at late enough times, primordial black holes of mass about 1 g could be substantially produced, saturating the observational
bounds when a Planck mass remnant survives the evaporation
process. Interestingly enough, a smaller value of the field coupling
$g^2 = 6.5 \times10^{-8}$ (case~I) shows no such overproduction of PBHs. This
is because of the development of a negatively skewed PDF as shown in
Fig.~\ref{fig:4}. 

One can argue that dependence of the PBH formation
rate on the coupling parameter $g^2$ is due to the existence of
resonance bands for the enhancement of fluctuations inside the
cosmological horizon \cite{Kofman:1997yn}, and 
some studies have argued that PBH overproduction is not a generic
feature of preheating. Indeed, Ref.~\cite{Suyama:2004mz} finds an
insignificant mass fraction $\beta_{\rm PBH}$ assuming matter
fluctuations collapsing at horizon scales and with a Gaussian
distribution of probability. Our results show that the Gaussian
distribution underestimates the true probability for case II (see
the first column of Table~\ref{table:1} for a $\beta_{\rm PBH}$
computed from the Gaussian PDF). If we relax the assumptions of
Ref.~\cite{Suyama:2004mz} and consider the possibility of PBH formation
below the Hubble scale, we find a clearly non-Gaussian PDF from the
distribution of fluctuations in the numerical simulation. Our
estimations of the PBH mass fraction in Table~\ref{table:1} indicate
that the overproduction of PBH is a likely feature in 
preheating, and this possibility must be studied in  more detail. 

In cosmology, PBHs present us with a unique tool to probe the small scales
which are not covered by CMB and large scale structure surveys.
In the present paper, we show that, while strong couplings
between the fields are useful to reheat the Universe efficiently, it
is important to test these viable models for PBH overproduction due to
the highly nonlinear physics involved in the preheating stage. In
this work, we have put forward a method to account for the mass
fraction of PBHs, and we shall eventually use it to constrain
preheating models.  

Motivated by the results presented here, we aim to evaluate a much
larger sample of coupling values $g^2$ in a subsequent work, where we
shall also take into account the evolution of metric perturbations and
explore a larger range of scales.  Another aspect that deserves
further study is the criterion used to determine which overdensities
will collapse to form PBHs during preheating. The threshold values reported in the
literature for scalar fields or barotropic fluids consider mostly
collapsing inhomogeneities as soon as they enter the cosmological
horizon. The growth of large inhomogeneities due to the nonlinear
interactions inside the horizon might eventually require a criterion
for PBH formation at subhorizon scales beyond that presented in
Ref.~\cite{Lyth:2005ze}, in order to constrain the parameter space with
even higher accuracy.

We finally note that the overproduction of PBHs could alter the
mechanism of transition to a radiation-dominated stage: It is known
that preheating in chaotic models cannot drive the Universe to a
radiation stage, and one has to consider other higher couplings in the
fields to achieve  that
\cite{Podolsky:2005bw,Dufaux:2006ee}. According to our results, 
the overproduction of PBHs during preheating opens the possibility to
reconsider PBH evaporation as an auxiliary mechanism to reheat the
Universe \cite{Hidalgo:2011fj}; such a mechanism may benefit
unification models \cite{Liddle:2006qz,*Liddle:2008bm}.

\begin{table}[t]
  \centering
\begin{tabular}{| l || c | c | c | c | }
  \hline    
  $ L_H  / L $   &   Case  & $  \beta_{\rm gauss}  $  &   $  \beta_{\rm max}  $   &   $  \beta_{\rm min} $   \\
  \hline\hline   
  $46 $ & I & $2.94\times 10^{-31} $  & $2.94\times 10^{-31} $ &  $ 6.05 \times 10^{-64}$   \\
  $46 $  & II &  $5.53\times 10^{-6} $  &   $ 4.7 \times 10^{-3}$ &  $ 1.5 \times 10^{-4}$   \\
     \hline    \hline 
     $ 23 $ & I & $4.59\times10^{-32} $  & $   4.59\times10^{-32} $ &  $5.83\times10^{-67} $  \\
  $23$  & II &  $2.26\times 10^{-16} $  &   $ 2.6 \times 10^{-4}$ &  $ 6.63 \times 10^{-10}$   \\ 
  \hline    \hline 
     $16 $ & I & $ 2.20\times10^{-33}$  & $2.20\times10^{-33}$ &
  $1.23\times 10^{-71}$   \\ 
  $16$  & II &  $1.37\times 10^{-28} $  &   $ 2.3 \times 10^{-5}$ &  $ 4.52 \times 10^{-16}$   \\ 
  \hline\hline   
  $9$ & I & $3.19\times 10^{-37}$  & $3.19\times 10^{-37}$ &  $ 7.93
  \times 10^{-87}$   \\ 
  $9$  & II &  $2.37\times 10^{-64} $  &   $ 2.6 \times 10^{-7}$ &  $ 5.49 \times 10^{-33}$   \\
  \hline\hline   
  $8$ & I & $5.14\times 10^{-40}$  & $5.14\times 10^{-40}$ &  $ 5.87 \times 10^{-94}$   \\
  $8$  & II &  $2.58\times 10^{-80} $  &   $ 5.32 \times 10^{-8}$ &  $ 3.98 \times 10^{-40}$   \\
 \hline\hline   
  $6$ & I & $2.38\times 10^{-43} $  & $2.38\times 10^{-43} $ &  $ 5.95
 \times 10^{-114}$   \\ 
  $6$  & II &  $7.18\times 10^{-99} $  &   $ 1.15 \times 10^{-8}$ &  $ 3.39 \times 10^{-49}$   \\
   \hline  
\end{tabular}
\caption{$\betapbhs$ estimations for modes inside the horizon in both
  cases I and II computed from Eq.~\eqref{mass:fraction} with a threshold
  value $\zeta_{\rm th} = 0.7$. The statistics of smaller modes
  presents a considerable uncertainty due to the resolution of our
  numerical simulation.}  
\label{table:1}
\end{table}


\begin{acknowledgments}
E.T.L. acknowledges support from CONACyT M\'exico and  would like to 
thank the kind hospitality of the QMUL.
 L.A.U.-L. was partially supported by PROMEP, DAIP, PIFI, and by CONACyT
M\'exico under Grant No. 167335, and No. 182445, the Fundaci\'on Marcos
Moshinsky, and the Instituto Avanzado de Cosmologia (IAC)
collaboration. J.C.H. acknowledges support from CONACYT (program
\emph{Estancias Posdoctorales y Sab\'aticas al Extranjero para la
  Consolidaci\'on de Grupos de Investigaci\'on}), and from Grants No.
 UNAM-PAPIIT-IA-101414-1 and No. UNAM-PAPIIT-IN103413-3. K.A.M. is supported, in part, by STFC Grant No.
ST/J001546/1. 
\end{acknowledgments}


\bibliography{PBH_Biblio}
\end{document}